\documentclass{article}
\begin{document}

               STATISTICAL MECHANICS OF  DYNAMICAL SYSTEMS

                   WITH TOPOLOGICAL PHASE TRANSITIONS

                                            AJAY PATWARDHAN

                              Physics Department, St Xavier's college, Mumbai

                            Visitor, Institute of Mathematical Sciences, Chennai

                                              ABSTRACT

Dynamical system properties give rise to effects in Statistical mechanics. Topological index changes can be the basis for phase transitions. The Euler characteristic  is a versatile topological invariant that can be evaluated for model systems. These recent developments in the foundations of Statistical Mechanics, that are giving new results, provide insight into the Statistical thermodynamics of small N systems; such as molecular and spin clusters. 

This paper uses  model systems to give a basis for redefining partition functions in classical statistical mechanics. It includes the properties of dynamical systems namely , KAM Torii, singular points and chaotic regions. The equipotential surfaces and the Morse and Euler index for it are defined. The conditions for the topology change in configuration space, and its effect on the partition function and the ensemble  average quantities is found. The justification for topological phase transitions and their thermodynamic interpretation are discussed.

1.INTRODUCTION

Since the classic work of Boltzmann, Gibbs , Poincare , and Birkhoff the relation between the axiomatic foundation of Statistical Mechanics and the properties of dynamical systems have been investigated by many persons.

The Ergodic hypothesis in phase space for small N number of particles has to incorporate the conclusions of the Kolmogorov- Arnold- Moser theorem. For dynamical systems that are hyperbolic, the Kolmogorov entropy has a role to play.

 In systems with both integrability and chaos on subspaces of phase space, the axioms for ergodicity and canonical ensemble partition functions need a fresh approach.

 For systems with symmetry and symmetry breaking, the ensemble averages are redefined. The singularities of the Hamiltonian vector field give rise to topological indices on phase space. Their role in the foundations of statistical mechanics is recently being explored by Cohen and others. References [1 to 16].

The work of Ruelle, Sinai , Leibowitz, Ford, Gallavotti, Cohen and others in making the framework of statistical mechanics consistent with the developments in classical and quantum mechanics of non linear dynamical systems has led to the resolution of some of the old questions and opened new possibilities for a more general Statistical mechanics.

To obtain a complete theory inclusive of all the effects is still a distant goal. In an earlier arxiv eprint the author had given modifications to the partition functions by introducing the Kolmogorov entropy, the Casimir invariants, and the Euler number for micro partitions. In this paper the topological phase transitions introduced by Cohen and others are developed into a framework for Statistical Mechanics of Dynamical systems.

For a number of model systems the singularities of the Hamiltonian vector field on phase space are evaluated to give a topological index , the Euler number. The topology changing transitions and their relation to statistical thermodynamics of phase transitions is the topic of some recent research.

It is expected that underlying these properties exists a structure of the theory that connects dynamical systems to statistical mechanics and thermodynamics. There is also the motivation to arrive at a theory for any N particles that for small N will be applicable for molecular clusters, nanoscale systems etc.

In this paper some conditions and their consequences for Euler number dependent partition functions are obtained. It is well known that lattice gas , binary alloy and Ising spin models can be mapped onto one another.

 It is also known that a number of model potentials lead to solvable systems that have interpretations in these contexts and applications in topics of condensed matter physics and other subjects. Hence any generic properties that can be found can lead to answers to a number of fundamental questions of Statistical Mechanics.

2. STATISTICAL MECHANICS AND DYNAMICAL SYSTEMS

The $ 6N $ dimensional phase space of suitably chosen  canonically congugate variables $ (q_i , p_i ) $, $ i = 1$ to $ 3N $ has  Hamiltonian functions $ {H} $ giving their dynamics. Consider Hamiltonians written as kinetic energy plus potential energy, with kinetic energy quadratic in momentum.

 The Gibbs density function $ exp(- \beta H ) $ can be integrated over phase space to obtain the canonical partition function. The momentum integral is separated out and done as a gaussian integral.

 The configuration space integral is written in terms of the constant potential energy surfaces (equipotentials); which are independent of momentum.

The micro partitions are the collection of subsets of these equipotential surfaces with isolated singular points in each subset. The singularities of the Hamiltonian vector field are the points at which the potential energy per particle has its partial derivatives equal to zero.

 The Hessian of the Hamiltonian is essentially reduced to the second partials of the potential energy function; and it is evaluated at the critical points. Its determinant gives the stability criteria and its eigenvalues give the index for the vector field.

 In Morse theory, the global toplological index is obtained in terms of the local properties of Hessian evaluated at the critical points. The number of eigenvalues with negative real parts is used to find the Morse index which is the upper bound for the Betti numbers of the sub manifolds of the potential energy (equal to constant) hypersurfaces. The alternating sum of the Morse indices is the Euler number , which is a topological invariant.

Any changes in the Hamiltonian and in the equipotential submanifolds of configuration space that alter the Euler number are topology changing transformations. The consequence for the partition function is that it changes the measure of integration in such a way that the dynamical system has a statistical mechanics that introduces topological phase transitions.

Consider that the phase space consists of isolated KAM Torii interspersed with chaotic regions in general. The Hessian evaluated on phase space in the KAM torii region does not have negative real parts for its eigenvalues. Nor for the chaotic regions. The configuration subspace is a submanifold of phase space with critical points at which the Hessian (reduced) can have some negative  eigenvalues.

 Hence the topological index is relevant at these points. As the canonical ensemble requires integration over all the equipotential hyper  surfaces; the location and  number of the critical points and the spectrum of the Hessian at these points keeps changing from surface to surface.

 Moreover the ensemble of Hamiltonians  are the deformations of the model Hamiltonian as the parameters vary within the allowed range. The ensemble average should therefore be generalised to include the topology changing transitions among the equipotential surfaces as the parameters and the energy is varied. There will be a contribution from the deforming and collapsing KAM Torii as well as the K entropy dependence of the chaotic region to the partition function. The latter have been discussed in papers by  Cohen et al, as well as by me.

 The invariant measure on phase space is a collection of measures on these subsets. It is ergodic on a subspace which excludes the KAM Torii and the singular points. The measure on the equipotential hypersurfaces, is written by factoring out the critical points where $\nabla V = 0 $. The partition function is written as

 $ Z(\beta) = (\pi/\beta)^{N/2}\int dv exp(-\beta v) \int d\sigma/|\nabla V |$

Consider a dynamical system described classically by a set of canonically congugate variables with a Hamiltonian. In a representation in which the kinetic energy is quadratic in momentum and the potential energy is independent of momentum

The  $ H (q_i , p_i ) = \sum_i p_i^2 + V(q_i) $ gives a canonical partition function  $ exp( -\beta H ) $.

 Since the equipotential surfaces $ V(q_i) = v $ span the configuration space for values of $ V $, these  sub space of phase space have critical points or singularities where the  $ \nabla V = 0 $. At these points the Hessian reduces to the second partial derivatives

 $ \partial^2 V(q_i,q_j)/\partial q_i \partial q_j$

 which can have a non zero determinant(non degenerate case). It has eigenvalues that could be distinct ( non degenerate case) with a number $ k $ of them having their real parts negative. A index can be defined with this number.

 The Euler index is the sum $ \chi(M) =  \sum_k (-1)^k \mu_k(M) $.

 The Morse index $ \mu_k(M) $ is the upper bound of the Betti numbers $ b_k(M) $. For a number of holes and handles in configuration space, the Euler number can be written as $ \chi = 2 -2g $ in terms of the genus $ g $ equal to number of holes minus number of handles minus number of inner boundaries around singular points.

Following the model examples published and the methods of calculation developed in the references, it is possible to address the following questions.

 1) Is a modification of the foundation of Statistical Mechanics required for the topological properties.

 2) Is there a new dynamics dependent property introduced by the topological index.

 3) If the Euler index changes will this be a phase transition for the system at the level of its dynamics, with consequences for the canonical ensemble average values.

 4) Do these consequences imply a thermodynamic effect , a phase transition of a usual kind ( 1st or 2nd order), or is this a novel type of phase transition.

 5) How can the imprint of the singularities of equipotentials be seen at each stage of the calculation leading to observable effects for thermal variables and involving the Euler index.

3. MODELS OF DYNAMICAL SYSTEMS AND TOPOLOGICAL EULER INDEX

A wide class of interesting potentials such as van Hove, Takahashi, Fisher, van der Waal with hard core,  are suitable to work out the framework of calculation for topological phase transitions. In many of these models  the potential is infinite upto some $0<r<a$ and is zero for some  $r>b$; with various forms of functions, constant, linear, quadratic etc in between $ a<r<b$ . For this paper the model chosen  features a hard core with a region around it with a finite potential which goes to zero after some finite range.

 So $ V(r)$ is undefined , infinite for $ r<a $ and zero for $ r>b$. It is  a function with a minimum in between $a<r<b $. For a system of N particles with positions $ r_i $ the potential function $V(r_i,r_j)$ is defined as the above function for the distance $ 2a<|r_i-r_j|<2b $; it is undefined for $|r_i - r_j|< 2a$ and is zero for $|r_i-r_j|>2b.$ The volume  $ 4/3\pi a^3 $ is excluded around each particle.

 The particles have a minimum separation $2a$ and for separation more than $2b$ they do not interact. This could be interpreted as a finite size effect and a short range interaction.The parameters $a$ and $ b $ could be adjusted. One more parameter $ c $ could be introduced for the location of the minimum of potential.

 The form of the function of the potential could be any polynomial, reciprocal polynomial, smooth function and function with no other singular points in this range $ a<r<b $.

 Using $ V(r_i) = \sum_j V(|r_i - r_j|) $ the critical points can be found for $ \partial V/\partial r_i  =0. $ The Hessian reduces to the second partials

 $\partial^2V/\partial r_i\partial r_j $; to be evaluated at the critical points.

 This model involves the N particles moving around with the configuration space consisting of N moving holes of radius $a$ with essentially ideal gas behaviour for mean particle separation greater than $2b$ and no clustering allowed for particle separation less than $ 2a $.

 Suppose the form of the function $ V(r)$ was fitted with a quadratic

 $ V(r) = a_0 + a_1 r + a_2 r^2 $ for $a<r<b$,

 then using the values of the function $V(r)$  at $r=a, r=b $ and its minimum at $r=c$ 

,and solving  simeltaneuosly for these three parameters $a_0,a_1,a_2$ in terms of  $a,b,c$; these  can be substituted in the formula for $ V(r) $. There are no further undetermined parameters.

 If the empirically obtained model potential function, to some accuracy, fits the quadratic formula then its parameters from data ,can be mapped onto the two sets of parameters $a_0,a_1,a_2$ and $a,b,c$. This characterises the model completely in a testable manner.

 Now in the Hessian condition a direct substitution will enable a check on the coefficient $a_2$ and hence on the sign condition for the determinant and the eigenvalues real part. For higher order polynomials and other functions there will be additional parameters that may need empirical or theoretical inputs. For condensed matter potential examples, such as Calogero Sutherland, the core and  range parameters $a$ and $b$ respectively could be adjusted to be very small and very large. 

The topological index , $\chi = 2-2g$ with genus $g$, here the number of configuration space holes are $N$ for the N particle system. When clusters of size $n$ particles are formed and inscribed in a sphere of maximum radius $ 2b n^1/3  $ there are inner boundaries formed in configuration space equal to number of clusters $ M$ say.

 The genus $g$ is then $ N-M $ and so the Euler number has a contribution $2-2(N-M)$ which is a very large  negative number at low densities  as $ M $ is small. Then $g$ can reduce as M can increase upto $N/2$ and $\chi -> - N$. So the per particle contribution to $\chi$ is $-1$.

 In addition the topological changes that the potential energy surfaces have will also change $\chi$. The potential energy per particle is taken to find the contribution to the Euler number from the singular points.

The Morse function is the number of negative coefficients of a quadratic form in the coordinates near the singular points. The equipotential surface is considered to be mapped into this quadratic

 $ -x_1^2 .....-x_k^2 + x_{k+1}^2 .....x_N^2 $ , locally at the singular points 

and the Hessian which has the second partial derivatives has therefore a set of real eigenvalues with $ k$ negative ones and  $ N-k$ non negative ones.

 In the model potential chosen the coefficients of the quadratic expression are fully  determined  for each coordinate and so the sign of the  eigen values  in the Hessian is fixed. This is diffeomorphism independent.

 In crossing the critical point the contribution is picked up to the Morse index; as a handle of index $ k $ in the configuration space. The index $ k $ and coindex $ N-k $ for the Morse function gives a permutation among the $ k $ and the $ N-k $; $ N!/k!(N-k)! $ and the number of handles $ k $ give the contribution to the genus $ g = N-M-k $ and hence to the Euler number $\chi = 2-2g $.

 Morse theory gives a relation between the number $k$  of negative eigenvalues and the number $ k $  of handles. The co index comes from the $N-k$ nonnegative eigenvalues. Hence the Morse index contains the permutations on the $N , k, (N-k)$ ; that is $ ^NC_k $. 

For the quadratic potential functions considered above, in a two dimensional model, there is a correspondence from the lattice gas to the spin system XY planar (Heisenberg) model as considered in many publications. References [1 to 16].

 $V(\phi) = J/2N \sum_{i,j}^N [1 - Cos( \phi_i -\phi_j )] -h \sum_i^N Cos (\phi_i)$ becomes

 $ V(m_x,m_y)  = JN/2(1-m_x^2 -m_y^2) -h N m_x $ ;

 $ J>0 $ is the coupling and units are chosen so that it is set equal to $1$,

 and  $ h>0 $ is the external field

 and hence a equipotential surface per particle $V/N = v $ can be defined.

 The variables $m_x , m_y $ are bounded by [-1,+1] :

 $ m_x = 1/N \sum_i^N Cos (\phi_i)$ and

 $ m_y = 1/N \sum_i^N Sin (\phi_i)$ 

In mean field theory the following are obtained:

For this potential the $\nabla V = 0 $ condition gives the following  :

$ tan(\phi) = m_y/(m_x + h)$

and the Hessian condition gives the following:

$( m_x + h) Cos(\phi) + m_y Sin (\phi)$

 as the diagonal part and the off diagonal part is

$- \sum_{i,j} Cos(\phi_i)Cos(\phi_j) + Sin(\phi_i)Sin(\phi_j).$

The determinant non zero condition is satisfied at the critical point as checked by substitution.
      
And the eigen values are found from the condition on independent $\phi$; such that the sign of $ m_x + h $ determines sign of $ tan(\phi)$ and  the number $ k $ of negative eigenvalues.

 This  $ k $ is bounded by a number $n$ such that $m_x = 1-n/(N/2)$

 gives a sign change even for $h --> 0$, when $ n>N/2 $ changes to $ n<N/2 $. 

This can be interpreted as the onset of clustering as dimers are formed. When the spin system model is mapped into the molecular lattice gas model the spin clusters are replaced by molecular clusters.
       
From these numbers the Morse index is found as follows: 

Replacing $ m_x $ and $ m_y $ by $( N-2n)/N $ in the equipotential formula

 gives $v(n) = 1/2 [ 1 - 1/N^2(N-2n)^2 ] -h/N (N -2n)$.

 This can be solved for $n$  as integer value of the quadratic solution

 $ n(v)= Int[ (1+h)N/2 +/- N/2 ( h^2 - 2 (v - 1/2)^{1/2}].$

 In this model the range of $v$, the potential energy per particle is bounded.

 With the $m_x$ and $m_y --> 1$; the $ v --> -h $ and 

with the discriminant in the quadratic solution required to be positive

 $ h^2 > 2(v - 1/2 )$ gives $ v < 1/2 + h^2/2$ .

 Hence the condition $ -h < v < 1/2(1 + h^2 ) $.

 The integral in the partition function is effectively on this range.

 The Morse index is $ \mu_k (v)  = N!/k!(N-k)![1-\theta (k-n(v)) + \theta (N-k-n(v))] $

 so that for $ k<N/2 $ it is $ ^NC_k $ and for $ k> N/2$ it is zero since there are $ k $ negative eigenvalues bounded by the $n(v)$.
       
The Euler number is given by the alternating sum :

 $ \chi(v) = \sum_k (-1)^k \mu_k(v)$. This varies with the equipotential chosen $v$.

 However in this model there is one transition point as the $v$ varies; with the $ n $ switching between its two values ; roots of quadratic.

 For example if $v--> -h$ then $ n = Int(Nh) $ or $ n = N $. $ k$ is upper  bound by $n$.

 Also as $ v --> 1/2 +h^2/2 $ then $n$ has one fixed value

 $ n = Int(1+h)N/2 $. 

If the coupling $ J $ had not been set equal to $1$ and retained in the calculation , then these conditions would depend on  $ J $ also.

Hence it is clear that at some in between value of $v$ in its allowed range ,the qualitative change occurs when $ n $ goes from single value to double value. This leads to the required change in the  Morse index and the Euler number.

 The potential energy ``landscape'' with its minima , maxima, saddle points is contributing through the local minima to the index. As the partition function integral goes over the equipotentials $ v $ , these indices are preserved except at the transition point. A topological change on configuration space leads to a phase transition , and it occurs at the corresponding  temperature.

To have a topological phase transition requires a change in the Euler number; for which the conditions are on the $ (-1)^{n(v)}(  ^NC_{n(v)}) $ as the $ n(v) $ switches from single value to double value; a phase transitionoccurs  between two phases. The diffeomorphism invariance of the equipotential hypersurfaces ,submanifolds in configuration space is broken at $v<v_c $ and $v>v_c$.

The relation to thermodynamic first and second order phase transitions is demonstrated in recent references.

 In the system of coupled anharmonic oscillators

 $ V = \sum_{i,j} \alpha_{ij} q_i q_j  + \sum_{i,j} \gamma_{ij} q_i^2 q_j^2$

 the Riemannian geometry on phase space is used to obtain geodesic and geodesic deviation equations. In chaotic regions the Lyapunov exponents and hence the  exp(Kolmogorov entropy) measure is defined in terms of the Riemann tensor.

 For the singular points and Morse function the second curvature form from Gauss Bonet theorem gives the Euler index. In the integrable regions the $[ I_i, I_j]= 0 $  Poisson Bracket for integrals in involution ; gives a measure

 $ exp( -\beta \sum_m \lambda_m I_m )$ with lagrange multipliers $\lambda_m $. 

This is defined on the KAM Torii of dimension $m$.

 Explicit computations of Transition temperature, Entropy and Free Energy are reported. However in comparison to the standard methods and results of statistical thermodynamics of phase transitions this work is at a preliminary stage. It has  been proved that there is a definite effect of the underlying dynamics in terms of the topological invariants on the canonical ensemble averages in classical statistical mechanics.

For quantum statistical mechanics the $ Tr(\rho) $ and $ Tr(\rho H )$ computation for $ \rho = exp(- \beta H )$can include the topological index on the configuration space by a) switching to path integral approach for partition function and b) adopting a Hilbert space of a composite sytem written in terms of the fundamental group representations.

 This subject is in a beginning stage for particle systems in condensed matter , although the work in field theory and string theory on partition functions with topology is substantial. The quantum spin chain is a prototype model which can be done in mean field theory, very similarly to the XY model; with expectation values replacing the $m_x$ and $m_y$. Then using the Action based partition function

 $Z = \int d[\phi] exp (-S[\phi])$ the calculations could be performed.

 This approach is at a formative stage in the subject of Topological phase transitions.

4.PARTITION FUNCTIONS AND TOPOLOGICAL PHASE TRANSITIONS IN STATISTICAL MECHANICS

For the model potentials chosen the partition fumctions are written as follows:

$ Z_N (\beta) = \int d(Nv) ( \omega_N(v) ) exp (- \beta Nv) $

 where $ v $ is the equipotential per particle, and $\omega$  is an additional measure , distinct in each region : KAM Torii, chaotic and singular; as described previously.

 For the KAM Torii it is $ exp(-\beta\sum_m \lambda_m I_m(v))$;

for the chaotic regions it is $ exp(Kolmogorov entropy(v) )$

 For the singular point contribution 

$ \omega_1(v) = \int d\sigma/||\nabla V || $ and 

 with $\omega_N(v) = (\omega_1(v))^N $

$ Z_N(\beta) = N \int d(v) exp (N ( ln \omega_1(v) -\beta v ))$ 

Hence the average potential  energy per particle is

$<v(\beta)> = 1/N [ - \partial ln Z_N(\beta)/\partial\beta ]$

This also indicates that the factor $N ln\omega_1(v)$ enters like an additional entropy in the exponent for the partition function integral.

The topological phase transition is induced by computing the $ 1/N ln(\chi(v))$ and plotting versus $ v $. Model systems calculcated in references give a sharp jump  discontinuity in the derivative of this quantity at a critical $ v= v_c$.

 For the example of the XY model above this behaviour can be traced to the factor $ ^NC_{n(v)} $ and what happens to it at the $v_c$. Deformations in the equipotentials induced by tuning the external field $ H $ or the parameters $J$ in the coupling could be the external and internal, respectively, causes for such a transition.

 As seen in the previous example the measure is modified by the Euler index and it enters into the exponent of the partition function integrand as an additional entropy. Entropy expressed as a logarithm of the alternating  sum of Morse indices or Betti numbers or genus numbers as the case may be.

It is therefore seen that the transition temperature $ T_c = 1/ k_B\beta_c $ is related to the $<v_c> $of critical potential energy per particle , thus directly relating the thermodynamic phase transition to the topological pphase transition. The average energy is computed and the entropy has a contribution from the Euler numbers.

 The Helmhotz Free energy as $F = -(\beta)^{-1}lnZ_N(\beta)$ will have its derivatives becoming discontinuous as the $v$ crosses $v_c$ and the single valued $n(v)$ become double valued. The correspondence with the usual description of phase transitions of 1st and 2nd partial derivatives of Free energy becoming discontinuous is being demonstrated by computing example systems ; and there are attempts to proove theorems.

 Many of the results obtained will remain valid for classes of Hamiltonians that have a infinite or very large potential for the  core and a finite range beyond which the potential is zero or very small; with at least a lower bound for the potential energy in between core and range, with one or more minima.

Example systems with multiple critical points , constraints, degenerate Hessians and model Hamiltonians often used in condensed matter physics are some open questions. Publications with complete studies of specific models are increasing recently. With Ising like , Potts, spherical, rotor chain etc models considered for nano and mesoscale systems; statistical thermodynamics of ``small'' and ``moderately large''systems. References[1 to 16].

5.CONCLUSION

This paper explores a growing area of foundations of statistical mechanics ; with possible application to small N systems too. The advances in Dynamics, Classical and Quantum, have led to new work in Statistical Mechanics. The mathematical properties of chaos and integrability, and topology and geometry as found in dynamical systems also creates  new possibilities for Statistical Mechanics. The definitions of Partition functions and the calculation of average values that includes the new properties, also give rise to new effects.

 This direction of work will give insights into the previously known Statistical Thermodynamics , as well as extend the subject to 'any N'; small and big  number of particle systems. It is therefore  a significant step in the completion of the program of Statistical Mechanics. Its thermodynamic implications in a variety of condensed matter physics phenomena will require further exploration.

In this paper an attempt was made to bring some elements of this ``paradigmatic `` shift into a few model systems that capture the essential requirements for Topological Phase Transitions in the Statistical Mechanics of Dynamical systems. Conditions were found for such a phase transition and its interpretation.

6.ACKNOWLEDGEMENTS

I would like to thank Prof R. Jagannathan for my visit to the Institute of Mathematical sciences, Chennai. The support of the Institute Director and staff for my visit is appreciated. The discussions with the Institute members ; particularly with Prof G. Baskaran, and Prof P. Ray are acknowledged.

7.REFERENCES

1. E G D Cohen , L Casetti, M.Pettini Physics Reports 337(2000),237

2.www.arxiv.org/cond-mat 0303200, 9912092, 9903418, 9810406, 0104267 E G D Cohen et al

3.R Palais, C Terng Critical point theory and submanifold geometry Springer lectures on maths vol 1988

4.C.Thompson Classical equilibrium statistical mechanics, Oxford science publications, Clarendon press 1988.

5.Caiani, Casetti,Clementi,Pettini PRL 79, (1997),4311

6.Frazosi,Casetti, Spinelli, Pettini Phys Rev E 60 (1999) R5009; cond-mat 9810180

7.Casetti, Cohen, Pettini PRL 82 (1999),4160

8.Franzosi, Pettini, Spinelli PRL 84(2000),2774

9.cond-mat 0405421 ,0406760,0410282 topology and phase transitions

10.Phys Rev E ,71 (2005),036152

11.M.Pettini,L. Casetti,L. Spinelli,R.Franzosi cond-mat 9911235,9904266, 0104110

12.R Franzosi,M Pettini,L Spinelli math-ph 0505057, 0505058

13.Angelani,Zamponi, Ruocco cond-mat 0506200

14.M.Kastner cond -mat 0412199,0509136,0404511

15.cond-mat 0205483,0508419,9904073

16.Ajay Patwardhan cond-mat 0411176

\end{document}